# An estimate of the average number of recessive lethal mutations carried by humans

**Short title:** Mean number of recessive lethal mutations in human

Ziyue Gao[a,1], Darrel Waggoner[b,c], Matthew Stephens[b,d], Carole Ober[b,e] and Molly Przeworski[f,g,1]

[a] Committee on Genetics, Genomics and Systems Biology, University of Chicago

[b] Department of Human Genetics, University of Chicago

[c] Department of Pediatrics, University of Chicago

[d] Department of Statistics, University of Chicago

[e] Department of Obstetrics and Gynecology, University of Chicago

[f] Department of Biological Sciences, Columbia University

[g] Department of Systems Biology, Columbia University

[1] Corresponding authors: Ziyue Gao, 920 E 58th St., CLSC 413, Chicago, 773-702-2750, ziyuegao@uchicago.edu; Molly Przeworski, 608A Fairchild Center, 1212 Amsterdam Ave., New York, 212-854-9063, mp3284@columbia.edu

**Abstract**

The effects of inbreeding on human health depend critically on the number and severity of recessive, deleterious mutations carried by individuals. In humans, existing estimates of these quantities are based on comparisons between consanguineous and non-consanguineous couples, an approach that confounds socioeconomic and genetic effects of inbreeding. To circumvent this limitation, we focused on a founder population with almost complete Mendelian disease ascertainment and a known pedigree. By considering all recessive lethal diseases reported in the pedigree and simulating allele transmissions, we estimated that each haploid set of human autosomes carries on average 0.29 (95% credible interval [0.10, 0.83]) autosomal, recessive alleles that lead to complete sterility or severe disorders at birth or before reproductive age when homozygous. Comparison to existing estimates of the deleterious effects of all recessive alleles suggests that a substantial fraction of the burden of autosomal, recessive variants is due to single mutations that lead to death between birth and reproductive age. In turn, the comparison to estimates from other eukaryotes points to a surprising constancy of the average number of recessive lethal mutations across organisms with markedly different genome sizes.

**Introduction**

In diploid organisms such as humans, the efficacy of selection on a mutation depends both on the fitness of homozygotes and the fitness of heterozygotes, which reflects dominance relationships among alleles. Since recently introduced mutations are mostly present in heterozygotes, they will be purged less effectively by selection when recessive, and segregate at higher frequencies compared to dominant or semi-dominant alleles that cause a similar fitness reduction in homozygotes. For this reason, recessive alleles are expected to constitute a large fraction of strongly deleterious alleles segregating in diploid populations, and in particular of Mendelian disease mutations in humans.

One context in which the effects of recessive mutations are unmasked is in the presence of inbreeding, which leads to an excess of homozygotes compared to Hardy-Weinberg expectation. Because closely related individuals may co-inherit alleles from one or more common ancestors, the genomes of offspring of consanguineous couples are more likely to be identical by descent, revealing recessive, deleterious traits. If there are many recessive or nearly recessive, deleterious mutations (i.e., mutations for which the fitness of the heterozygote is close to the fitter homozygote) segregating in the population, inbred individuals will, on average, have lower fitness than outbred individuals. A reduction in mean fitness due to inbreeding ("inbreeding depression") has been demonstrated repeatedly in experimental studies in multiple *Drosophila* species (e.g., (1); reviewed in (2)), as well as under more natural conditions in mice (3). In humans, genetic effects of inbreeding remain poorly quantified, because consanguinity is usually associated with non-genetic factors such as socio-economic status and maternal age, and genetic and non-genetic factors are hard to disentangle. Estimating the burden of recessive deleterious mutations in humans is therefore key to predicting adverse outcomes of consanguineous unions due to genetic factors alone (4-7).

Two main methods have been developed to these ends. Both aim to quantify the deleterious health effects that arise in offspring of consanguineous matings. The first considers couples with variable degrees of relatedness, regressing the viabilities of their offspring on their inbreeding coefficients, $F$ (4). When applied to humans, this method is subject to a number of important limitations. For one, the estimate relies heavily on accurate assessments of degrees of relatedness, and yet the $F$ values estimated from recent pedigrees do not capture inbreeding among more distant ancestors. This will bias the results if, as seems plausible, consanguineous marriages tend to occur in families with a tradition of close-kin unions (8). Moreover, in practice, due to the restricted range of $F$ and the small number of data points, the quantitative estimate of the combined effect of recessive deleterious mutations is highly sensitive to the choice of the regression model (9). Perhaps most importantly, consanguineous and non-consanguineous groups likely differ with respect to socioeconomic factors, in ways that influence the mortality and morbidity of the progeny. Depending on the direction of these effects, this could lead to either overestimation or underestimation of the genetic effects of consanguineous marriage on health outcomes.

To minimize these concerns, a second approach focuses specifically on the comparison between offspring of first-cousins marriages, the most common form of consanguineous unions, and of non-consanguineous marriages, in a large number of populations (5-7). Regression of the mortality of first-cousin progeny—for which the expected $F$ is 1/16—on that of non-consanguineous progeny in the same population reveals a significant excess mortality in the former, which is then translated into the aggregate effect of recessive deleterious mutations. Even in this approach, however, genetic effects may be confounded by socioeconomic conditions that differ between consanguineous and non-consanguineous groups *within* a population. Moreover, depending on the population, consanguinity levels may either increase or decrease with socioeconomic status, and thus using the excess mortality of first-cousin offspring could either under-estimate or over-estimate genetic effects.

Here, we introduce an approach that is not confounded by environmental effects, considering a founder population with a known pedigree and complete disease ascertainment over the past few generations. Founder populations have contributed greatly to the identification of Mendelian-disease mutations, because the founding bottleneck and subsequent inbreeding increase the chance of recent identity by descent and thus the incidence of a number of otherwise rare, recessive diseases (10). With a known pedigree, we can estimate the probability that an autosomal recessive, deleterious founder mutation manifests itself (i.e., occurs as homozygous in at least one individual) in the past few generations, by simulating its transmission down the pedigree. From this estimate and the number of recessive diseases observed in the pedigree, we can obtain an estimate of the total number of deleterious mutations carried by the founders. Since the number of founders is known, this translates into the average number of recessive lethal alleles in each haploid set of autosomes. An advantage of our approach is that, by directly utilizing the pedigree information, there is no need to calculate an inbreeding coefficient or to compare among groups that are potentially subject to different socioeconomic conditions. A difficulty, however, is that the transmission probability of a recessive deleterious allele depends on its selection coefficient in homozygotes ($s$), which is in general very hard to quantify. We therefore focused on autosomal, recessive lethal mutations ($s=1$), defined as mutations that when homozygous lead to complete sterility or death between birth and reproductive age, in absence of medical treatment. Since recessive lethal mutations are only a subset of all deleterious mutations, our estimate provides a lower bound on the burden of recessive deleterious mutations, as well as information on the tail of the distribution of fitness effects of deleterious mutations (e.g., 11, 12, 13).

**Results**

We focused on the Hutterites, a group ideally suited for this purpose. The Hutterian Brethren is an isolated founder population, which originated in the Tyrolean Alps in 1520s and eventually settled in North America on three communal farms in the 1870s after a series of migrations throughout Europe. The three colonies thrived and shortly thereafter gave rise to three major subdivisions, referred to as the Schmiedeleut (S-leut), Lehrerleut (L-leut) and Dariusleut (D-leut), with most marriages since 1910 taking place among individuals within the same leut. The Hutterites have kept extensive genealogical records, from which highly reliable pedigrees have been reconstructed (14-16). Moreover, researchers and the Hutterites community have built a close partnership, greatly facilitating the diagnosis and identification of genetic disorders. Incidences of genetic disorders in Hutterites have been comprehensively documented since late 1950s, with more than 40 Mendelian disorders, of which 35 are autosomal recessive, described in the literature (10, 17-22). The pedigree information and nearly complete ascertainment of genetic disorders over this time period make it possible to reliably infer the number of recessive lethal mutations in the founders.

Specifically, our analysis was based on a 13-generation pedigree that relates 1,642 extant S-leut Hutterites in South Dakota and their ancestors (3,657 individuals in total), all of whom can be traced back to 64 founders who lived in the early 18$^{th}$ to early 19$^{th}$ centuries in Europe (23). We ran gene-dropping simulations (similar to 23, 24) on the pedigree to assess the probability of a recessive lethal mutation manifesting itself, conditional on being carried by a founder on an autosome (see Methods for simulation procedures). We defined "manifestation" as the presence of at least one individual in the pedigree who was born after 1950— the period after which disease ascertainment is close to complete—and inherited two copies of a recessive lethal variant introduced by the founders.

We found that on average 57% of unique recessive lethal mutations carried by the founders will have been lost before 1950 (i.e., none of the individuals in the pedigree born after 1950 carried the mutation). This proportion is almost the same as for neutral variants with the same initial frequency in the founders, suggesting that the loss of variants is primarily due to the severe genetic drift after the founder event (confirming results found in (23)). Among the recessive lethal mutations that survive until 1950, we expect that 19.2% will have been exposed in homozogote(s), and thus overall, in expectation, 8.2% of all recessive lethal mutations carried by the founders will have manifested themselves to date. The probability is almost the same if we considered manifestation as the presence of homozygotes in individual(s) born after 1940, in order to account for a delay in diagnosis for diseases with an onset in adolescence (see Methods).

The above simulation scheme implicitly assumes complete reproductive compensation (see Methods), which might not be appropriate for all recessive lethal diseases. To address this concern, we ran a second set of simulations on a larger pedigree of 15,235 Hutterites, who can be traced back to 78 ancestors (who were not necessarily the "minimum ancestors" defined in (25); see Methods for details about this pedigree). The second simulation scheme does not make an assumption about reproductive compensation and would be exact on a complete pedigree, but it is potentially sensitive to the incompleteness of the pedigree (especially missing data in the early generations). Nonetheless, the results are similar: on average, 69% of recessive lethal mutations carried by the founders will have been lost before 1950 and 8.2% will have manifested themselves in individuals born after 1950.

Next, we considered all known autosomal recessive lethal diseases observed in the S-leut Hutterites included in the pedigree. To this end, we compiled a list of all known autosomal recessive disorders (11 in total) reported in S-leut Hutterites in the United States (see Methods for details). We classified each of the identified diseases as "lethal" if, untreated, it causes death prior to reproductive age or precludes reproduction for affected individuals. We found four such recessive lethal diseases: cystic fibrosis, non-syndromic mental retardation, a severe form of myopathy and restrictive dermopathy. The underlying mutations are known for all four diseases, and genotyping data of the 1,642 extant individuals confirmed the presence of homozygote(s) of the disease-causing mutations for the first three diseases (23). Restrictive dermopathy was excluded from our list when we used the 13-generation pedigree, because the only known case among the South Dakota Hutterites was not included the 3,657 individuals.

From the number of recessive lethal diseases (three) and the probability that a recessive lethal allele manifested itself since 1950 (0.082), we estimated that the total number of autosomal, recessive lethal mutations carried by the 64 founders is 3/0.082 = 36.6, or an average of 0.29 recessive lethal alleles in each haploid human genome (Figure 1A). To assess the uncertainty in this estimate, we estimated the posterior distribution of the mean number of mutations per haploid human genome conditional on observing exactly three diseases since 1950 (Methods and Fig 1B). If a uniform prior distribution is used, the posterior distribution has a mode of 0.29, and a 95% credible interval (CI) of [0.10, 0.83]. We also considered a uniform prior on the logarithmic scale in order to account for uncertainty in the order of magnitude, and a similar 95% CI is obtained (i.e., [0.059, 0.69] mutations per haploid genome). If we instead used simulation results from the larger pedigree, the point estimate and 95% CI are similar (see Methods).

Simulations further indicate that only a small fraction of the surviving recessive lethal mutations have been seen in homozygotes, so there are more hidden, recessive lethal mutations that are segregating among extant individuals in the pedigree. In fact, carrier screening has identified heterozygotes for three more recessive lethal mutations in the extant S-leut Hutterites in South Dakota, which have manifested themselves in Hutterites outside the pedigree under study (Table S1) (23). Based on our simulation results, we expect as many as a dozen more such mutations in addition to these cases.

In generalizing from the results for the Hutterites to other human populations, one concern might be that their demographic history prior to the founder event in the 18-19[th] centuries was atypical in ways that influence the number of recessive lethals carried by the founders. In particular, a long period of endogamy could have purged recessive deleterious alleles from the population (26). This seems unlikely to have had a major effect, however: over the fifteen or so generations between the origin of the Hutterites in the 1520s and the founding event, even relatively high levels of human inbreeding ($F$=0.03) should only decrease the mean allele frequency of recessive lethals by ~30% (27). Moreover, such a decrease would be lessened or nullified by reproductive compensation (27), as might occur in the Hutterites (28). These considerations suggest that estimates based on the Hutterites should be broadly applicable and would, if anything, be slight lower than the mean number of recessive lethals carried by larger, outbred populations.

In that regard, we note that our estimate of the average number of recessive lethal mutations per haploid genome is lower than the previous estimates of the total number of lethal equivalents per haploid genome (0.56-0.7 in (6, 7)). A lethal equivalent is defined as a locus (or a set of loci) that, when in homozygous state, would cause on average one death, e.g., one lethal mutation or two mutations each with 50% probability of causing death (4). In other words, the total number of lethal equivalents in a haploid genome can be thought of as the sum of the deleterious effects of all recessive mutations carried by an individual. As expected then, our results suggest that recessive lethal mutations are only a subset of the recessive mutational burden. Interestingly, however, the difference between our point estimate and previous estimates is only about two-fold; even if we consider the lower bound of our credible interval on the mean number of recessive lethals, it is still about one sixth of the total number of lethal equivalents. Thus, a substantial portion of the burden of recessive mutations carried by humans may be attributable to mutations that lead to death between birth and reproductive age.

**Discussion**

Our approach indicates that on average, one in every two humans carries a recessive lethal allele on the autosomes that lead to lethality after birth and before reproductive age or to complete sterility. This estimate should be unaffected by socio-economic factors. Moreover, incomplete ascertainment of diseases is likely not a major concern, because most severe genetic disorders that occurred in Hutterites after the 1950s are expected to have been documented (10), so we expect at most a slight under-estimate of their number. In addition, while we ignore linkage between recessive lethal alleles when simulating the transmission of each mutation independently and linkage might lead to a greater variance in the proportion of mutations that survive or manifest themselves, it should not influence the mean proportion, so will not bias our estimate of the probability of manifestation obtained from simulations.

Beyond the Hutterites, this approach can be applied to other isolated founder populations with limited immigration, for which there is reliable genealogical information since the founding and close to complete disease phenotyping in the relatively recent past, such as the Amish (29) and the inhabitants of Norfolk island (30).

An important caveat, however, is that estimates from our approach are limited to lethal diseases that manifest themselves at or after birth. This issue is common to most studies that estimate the mutational burden in humans, because of the limited availability and reliability of data on prenatal loss. Studies that considered data on the frequency of miscarriages (i.e., a gestation age of 28 weeks or more) reported no or little effect of consanguinity on prenatal losses, while finding clear-cut effects on postnatal mortality (5, 31, 32). This cannot be taken as strong evidence for the absence of embryonic recessive lethal mutations in humans, however, as most losses due to recessive lethals may occur during the early stages of pregnancy. Even if the data on early pregnancy loss were available, the high rate of spontaneous pregnancy failure due to other causes (33) may obscure the difference between consanguineous and non-consanguineous groups due to genetic factors. In contrast to how little is known in humans, extensive mutation screens in mice reveal a high proportion (40-50%) of autosomal knock-out mutations that cause deaths in prenatal stages when homozygous (34, 35). If the proportion of embryonic lethals were similar for spontaneous mutations in humans, then each human individual would carry approximately one recessive lethal mutation that acts across ontogenesis (using the point estimate from our analysis).

**Some implications**. If we take our point estimate of the number of recessive lethal mutations per individual at face value, it suggests that the risk of autosomal recessive lethal disorders that manifest after birth should be increased by 0.29/16=1.8% in offspring of first cousin couples (assuming no difference in environmental factors). This prediction agrees well with the estimated 3.5-4.4% increased risk for pre-reproductive mortality and 1.7-2.8% increased incidence of congenital anomalies in children of first cousins above the general population risk (8).

Our results also provide insight into the number of autosomal sites in the human genome at which mutations would lead to lethality when in a homozygote state. The mean frequency of a recessive lethal allele in a finite population can be estimated under the assumption that Mendelian diseases reflect a balance between mutation, selection and drift (36). Assuming a random-mating, diploid population with constant effective population size of 10,000, a mutation rate of $1.2 \times 10^{-8}$ per bp per generation and given the estimate of 0.29 recessive lethal mutations per haploid set of autosomes, we predict that there should be ~85,000 autosomal base pairs at which mutations lead to recessive lethal disorders on or after birth (see SI). Recessive disease-causing mutations may not be completely recessive, however, in that carriers of one copy may also have a slight decrease in their fitness that is too subtle to be detectable in clinical diagnosis. If so, the mutations will segregate in the population at much lower frequencies due to selection against heterozygotes, and the target size could be much larger. For instance, if the heterozygous effect is a 1% decrease in fitness, the corresponding target size would become approximately 240,000 base pairs (derived assuming mutation-selection balance (37); see SI). While this estimate should not be taken too literally, as many recessive disease mutations are not point mutations (e.g., 38), it provides a sense of the minimal number of sites of critical functional importance in the human genome.

Moreover, this estimate of target size provides complementary information to population genetic approaches that aim to estimate the distribution of fitness effects of new mutations from polymorphism and divergence and mostly learn about weaker selection coefficients (39). These methods find that 25-40% of all amino acid changes in humans are strongly deleterious (i.e., have $s$>1% in a genic selection model) (11-13). Combining these estimates with our estimated target size would then suggest that 0.5-2% of strongly deleterious mutations are recessive lethals that are fatal between birth and reproductive age.

**Comparison to other species**. Intriguingly, our estimate of the average number of recessive lethal mutations per individual is in good agreement with what has been determined experimentally in a number of other diploid animal species. Most studies were conducted in *Drosophila melanogaster*, where individuals from natural or laboratory populations were made homozygous in order to measure the effects on viability. The results are relatively consistent, with on average 24.7% of the second chromosomes and 40.7% of the third chromosomes in the population carrying at least one recessive lethal (or nearly lethal) mutation (2). Assuming the number of such mutations is Poisson-distributed, this implies that each *D. melanogaster* harbors on average ~1.6 autosomal recessive lethal mutations. Similar numbers were obtained in other *Drosophila* species (e.g., (40, 41)), as well as by sibcrosses in *Lucania goodie* (1.87 lethal mutations per individual) and *Danio rerio* (1.43 mutations) (42).

Our estimate in humans (of approximately one recessive lethal mutation per individual that act across all developmental stages) is again quite similar, raising the question of why such distantly related organisms carry similar numbers of recessive lethal mutations per genome, despite their highly variable genome sizes (42, 43). Under a model of mutation-selection balance, the equilibrium frequency of a recessive lethal allele in an outbred population depends solely on the mutation rate at that site (44). Given that the mutation rates per base pair per generation are thought to be similar across vertebrates and *Drosophila* (45), organisms with larger genome sizes should therefore carry more recessive lethal alleles. One explanation for the finding in humans could be the mating patterns. Over the past centuries, consanguineous marriages were common practice, and the custom remains prevalent in many countries (8, 32). In the long term, depending on the degree of reproductive compensation (27), inbreeding can facilitate the purging of recessive deleterious mutations from the population (26), leading to a lower equilibrium mean frequency of recessive lethal mutations in humans compared more outbred organisms. An alternative explanation is that the number of recessive lethal mutations per individual reflects the total number of sites of critical, functional importance, which may be less variable across taxa than is the genome size or the number of genes.

**Methods**

**Simulations**

To assess the probability of a recessive lethal mutation manifesting itself after 1950s, we ran two sets of gene-dropping simulations:

(i) We first considered a "minimal pedigree", consisting of only those individuals ancestral to the 1,642 Hutterites for whom the genotypes at 14 autosomal recessive diseases-causing genes were determined (23). When using this pedigree, we assume no transmission distortion (46) and complete reproductive compensation, i.e., we assume that there is no relationship between the number of surviving children and whether the parents are carriers of recessive lethal diseases (27). In each replicate, we assigned a mutation to one founder at random and then simulated the genotypes of all other individuals generation by generation. For any individual in the minimal pedigree with children (either in the minimal pedigree itself or the larger pedigree), if both parents are carriers of a recessive lethal, we assigned a heterozygous genotype with probability 1/3 and a homozygous wild-type genotype with probability 2/3 (23). For individuals without known children, we simulated their genotype given the genotypes of the parents, according to Mendelian inheritance rules. This simulation scheme relies on the assumption of complete reproductive compensation, but is robust to missing data in the pedigree. After generating genotypes of all individuals in the pedigree, we examined the numbers of heterozygous or homozygous individuals for that particular mutation among people born after 1950. For each replicate, the mutation was classified as "lost" if there were no heterozygote or homozygote in the cohort, and as "manifested" if one or more homozygotes were among the cohort. Because the individuals in the pedigree are related, the incidences of a particular disease in the pedigree are not independent, so we considered the number of unique diseases instead of the number of instances. If reproductive compensation is not complete, this simulation scheme is expected to lead to an underestimate of the fraction of recessive lethal alleles that are purged since the founding and an overestimate of the probability of manifestation. We performed 64,000 gene-dropping simulations (1,000 replicates for each founder). The mutation was lost before 1950 in 36,529 (57.1%) cases and manifested in 5,259 (8.22%) cases. To account for a delay in diagnosis for diseases with an onset in adolescence or early adulthood, we also considered all individual(s) born after 1940 as the cohort. We performed 1,000 simulations for each founder, among which, the mutation was lost before 1940 in 36,227 (56.6%) cases and manifested in 5,337 (8.34%) cases. This scenario is the one that we focus on in the main text.

(ii) We also considered a larger pedigree, which consists of 15,236 individuals, all of whom can be traced back to 78 ancestors (who were not necessarily the "minimum ancestors" defined in (25)). Individuals who fell into one of the following three groups were included in this pedigree: (1) before the separation of the three leuts, individuals who had descendents who were S-leut; 2) all S-leut individuals who were born since the separation of the three leuts through 1980; 3) S-leut individuals who were born between 1980 and 2013 and participated in our ongoing studies (23). Using this larger pedigree, we make no assumption about reproductive compensation (but do assume that there is no transmission distortion (46)).

Specifically, in each replicate, we assigned a mutation to one founder at random and then simulated the genotypes of all other individuals generation by generation according to Mendelian inheritance rules. Because individuals homozygous for a recessive lethal mutation cannot reproduce, any individual who has offspring cannot be a homozygote. To model the transmission of recessive lethal mutations, we retained only replicates that are consistent with this condition, i.e., replicates where all individuals with offspring were either heterozygous or homozygous for the beneficial allele. This simulation scheme would be exact on a complete pedigree, but is sensitive to incompleteness of the pedigree (especially missing data in the early generations) in ways that are expected to lead to an overestimate of the fraction of recessive lethal alleles that are purged since the founding and an under-estimate of the probability of manifestation. We performed 78,000 gene-dropping simulations (1,000 replicates for each founder). Among the 67,964 replicates retained, the mutation was lost before 1950 in 46,719 (68.7%) cases and manifested in 5,563 (8.19%) cases. The proportion of mutations lost is higher than the one for neutral variants (60.0%), suggesting that in addition to the dominant effect of genetic drift, selection could also have played a role in purging out recessive lethal mutations in Hutterites. To account for a delay in diagnosis for diseases with an onset in adolescence or early adulthood, we also considered all individual(s) born after 1940 as the cohort. 500 replications were run for each founder. Among the 34,001 replicates retained, the mutation was lost before 1940 in 23,242 (68.4%) cases and manifested in 2,787 (8.20%) cases.

For both scenarios (i) and (ii), we also considered the situation where more than one copy of the same mutation was present in the founders. In each replicate, we randomly sampled two (or three) founders to be the carriers and simulated the genotypes of other individuals as described above. 100,000 simulations were run with the "minimum pedigree" for the case with two and three carriers, and the mutation was manifested in 16.7% and 25.5% of the cases, respectively. 10,000 simulations were performed with the larger pedigree with two and three carriers, and the mutation was manifested in 14.8% and 20.6% of the cases retained, respectively. These findings indicate that the probability of manifestation is approximately proportional to the number of carriers among the founders, enabling us to estimate the total number of recessive lethal alleles (some of which could be copies of the same mutation) carried by the founders by dividing the number of distinct diseases by the probability of manifestation obtained when there was only one carrier.

**Identification of recessive lethal diseases in Hutterites in the pedigree**

Most of the genetic diseases reported in Hutterites were summarized in a review by Boycott et al (10), and the list of recessive diseases was further updated by Chong et al. (23). To incorporate newly identified diseases, we searched PubMed for genetic diseases in Hutterites that were reported since (23). We also searched for diseases with terms "recessive" and "Hutterites" on the Online Mendelian Inheritance in Man (OMIM, an online catalogue of human genetic disorders and underlying genes) and confirmed that all entries are included in our list.

We then classified a disease as "recessive lethal" if (i) it has 100% penetrance in homozygotes; and (ii) the heterozygotes are asymptomatic (although they may still have subtly decreased fitness; see main text); and (iii) the disease leads to pre-reproductive lethality (e.g., restrictive dermopathy, cystic fibrosis in females) or complete sterility (e.g., cystic fibrosis in males) in the absence of treatment. We included infertility due to biological reasons (e.g., cystic fibrosis in males) or inability to reproduce if the phenotype of the condition leads to a reproductive fitness of zero due to social barriers (e.g., non-syndromic mental retardation, myopathy with movement disorder and intellectual disability).

To restrict the number of recessive lethal diseases to the pedigree under study, we required there to be affected individuals in the S-leut in South Dakota. This narrowed down the list to four diseases (cystic fibrosis, non-syndromic mental retardation, restrictive dermopathy and myopathy with movement disorder and intellectual disability). We excluded restrictive dermopathy when considering the "minimum pedigree", because the only reported patient in S-leut was not included in the minimum pedigree (although the parents are included in the minimum pedigree and confirmed to be carriers by genotyping) (23, 47). For the other three diseases, genotype data of the 1,642 extant individuals confirmed the presence of individual(s) homozygous for the disease allele in the pedigree (23).

We note that two *CFTR* mutations have been identified in the Hutterites. Both alleles lead to severe phenotype such that homozygous or compound heterozygous males are completely sterile, and homozygous or compound heterozygous females cannot survive to reproductive age in the absence of treatment. We therefore treat them as two copies of the same recessive lethal mutation. We further note that although the p.F508del mutation is common in Europeans, it is present on a single haplotype in Hutterites, suggesting that it was introduced into the population by only one founder (23). The p.M1101K mutation is rare in Europeans but was identified on two haplotypes in Hutterites (48). The two haplotypes differ at multiple polymorphic sites on both sides of the mutation, indicating either that at least two recombination events have occurred in this region or that the p.M1101K mutation was introduce by two founders (23). Therefore, it is likely that two or three carriers of these two *CFTR* mutations were present in the founders. Given that the probability of manifesting a mutation is approximately proportional to the number of carriers in the founders, we can treat it as introduced by only one founder.

**Point estimation and credible intervals (CI) for the mean number of mutations per haploid human genome**

We used a Bayesian approach to estimate the credible interval for mean number of recessive lethal alleles carried by each haploid set of human autosomes, $R$. Given that $D$ recessive lethal diseases have been observed, the posterior probability of $R$ is:

$$f_{R|D}(r) = \frac{f_R(r)P(D|R=r)}{\int_{r=0}^{\infty} f_R(r)P(D|R=r)dr} \propto f_R(r)P(D|R=r), \quad (1)$$

where $f_{R|D}(r)$ is the conditional probability of observing $D$ diseases, and $f_R(r)$ is the prior on $R$.

Let $X_i$ be the number of unique recessive lethal mutations carried by the $i^{\text{th}}$ founder, among which $Y_i$ mutations manifested themselves in the pedigree. We assume that each $X_i$ is independently Poisson-distributed:

$X_i \mid R=r \sim Poisson(2r).$    (2)

For simplicity, we assume all mutations carried by the founders are unique. If the transmission of each mutation is independent and each of the $X_i$ mutations carried by the $i^{\text{th}}$ founder has a manifestation probability of $p_i$, then $Y_i$ follows a binomial distribution:

$Y_i \mid X_i \sim Bin(X_i, p_i).$    (3)

Because of the thinning property of the Poisson distribution, conditional on $R$, $Y_i$ follows a Poisson distribution:

$Y_i \mid R=r \sim Poisson(2rp_i).$

Moreover, because each $Y_i$ only depends on the corresponding $X_i$ and, conditional on $R$, the $X_i$s are independent of each other, the $Y_i$s are independent Poisson random variables. Therefore, conditional on $R$, the total number of diseases observed, $D=Y_1+Y_2+\ldots+Y_{N_f}$, is a Poisson random variable:

$D \mid R=r \sim Poisson(2rpN_f),$    (4)

where $p=(p_1+p_2+\ldots+p_{N_f})/N_f$ represents the probability of manifestation if each mutation was equally likely to be carried by each founder, which can be estimated from our simulations.

Combining (1) and (4), we can re-write the posterior probability of $R$ as:

$$f_{R|D}(r) = \frac{f_R(r)\dfrac{(2rpN_f)^D}{D!}e^{-2rpN_f}}{\int_{r=0}^{\infty}f_R(r)\dfrac{(2rpN_f)^D}{D!}e^{-2rpN_f}dr} \propto f_R(r)\frac{(2rpN_f)^D}{D!}e^{-2rpN_f}.$$

Because there is little definitive information about $R$, we use a uniform prior on $(0, \infty)$:

$$f_R(r) = 1,$$

as well as a uniform prior of the logarithm of $R$ on $(0, \infty)$:

$$f_R(r) = \frac{1}{r}.$$

While both are improper priors, the resulting posterior distributions are proper, because the denominators of equation (4) converge in both cases. The mode can be solved as $R=D/2pN_f$, assuming a uniform prior (or $R=(D-1)/2pN_f$, assuming a uniform prior on the logarithmic scale). Given the value of $p$ obtained from the simulations, the 95% CI of the posterior distribution was solved numerically using *Mathematica* (49). For simulations with the "minimum pedigree" ($D=3$, $p=0.0822$, $N_f=64$), the CI for $R$ is [0.10, 0.83] when using a uniform prior and [0.059, 0.69] using a uniform prior on the logarithmic scale. Similar results were obtained from the larger pedigree ($D=4$, $p=0.0819$, $N_f=78$): the CI for $R$ is [0.13, 0.80] when using a uniform prior and [0.085, 0.69] using a uniform prior on the logarithmic scale.

**Acknowledgements**

We thank Jessica Chong for helpful comments regarding the pedigree data and simulations; Mark Abney for discussing the simulation procedures; Kym Boycott and Micheil Innes for their help in clarifying which recessive lethal diseases were present in S-leut Hutterites; and Dick Hudson, Marty Kreitman and Guy Sella for helpful discussions. This work was mostly done while M. P. was a Howard Hughes Medical Institute Early Career Scientist at the University of Chicago.

**Table 1. Three autosomal recessive lethal diseases and corresponding mutations observed in the smaller pedigree.**

| Name of Disease | OMIM Phenotype # | Gene | Mutation | Carrier freq. in Europeans | Carrier freq. in S-leut Hutterites (23) | Ref. |
|---|---|---|---|---|---|---|
| Cystic fibrosis[*] | 219700 | CFTR | p.F508del | 2.2-3.9% | 2.2% [†] | (48, 50) |
| | | | p.M1101K [a] | Unknown (one case reported) | 7.3% [†] | |
| Non-syndromic mental retardation | 614020 | TECR | p.P182L | Found only in Hutterites to date | 6.9% [†] | (18) |
| Myopathy with movement disorder and intellectual disability | -- | TRAPPC11 | c.1287+5G>A | Found only in Hutterites to date | 7% | (20) |

[*] See Methods for treatment of the two mutations in *CFTR*;

[†] The allele frequencies are reported in Chong et al., 2012 (23).

**References**

1. Greenberg R & Crow JF (1960) A Comparison of the Effect of Lethal and Detrimental Chromosomes from Drosophila Populations. *Genetics* 45(8):1153-1168.
2. Simmons MJ & Crow JF (1977) Mutations affecting fitness in Drosophila populations. *Annu Rev Genet* 11:49-78.
3. Meagher S, Penn DJ, & Potts WK (2000) Male-male competition magnifies inbreeding depression in wild house mice. *Proc Natl Acad Sci U S A* 97(7):3324-3329.
4. Morton NE, Crow JF, & Muller HJ (1956) An estimate of the mutational damage in man from data on consanguineous marriages. *Proc Natl Acad Sci U S A* 42(11):855-863.
5. Bittles AH & Makov E (1988) Inbreeding in human populations: an assessment of the costs. *Human mating patterns*, eds Mascie-Taylor CGN & Boyce AJ (Cambridge University Press, Cambridge England ; New York), pp 153-167.
6. Bittles AH & Neel JV (1994) The costs of human inbreeding and their implications for variations at the DNA level. *Nat Genet* 8(2):117-121.
7. Bittles AH & Black ML (2010) Consanguinity, human evolution, and complex diseases. (Proc Natl Acad Sci U S A), pp 1779–1786.
8. Hamamy H, *et al.* (2011) Consanguineous marriages, pearls and perils: Geneva International Consanguinity Workshop Report. *Genet Med* 13(9):841-847.
9. Makov E & Bittles AH (1986) On the choice of mathematical models for the estimation of lethal gene equivalents in man. *Heredity (Edinb)* 57 ( Pt 3):377-380.
10. Boycott KM, *et al.* (2008) Clinical genetics and the Hutterite population: a review of Mendelian disorders. *Am J Med Genet A* 146A(8):1088-1098.
11. Yampolsky LY, Kondrashov FA, & Kondrashov AS (2005) Distribution of the strength of selection against amino acid replacements in human proteins. *Hum Mol Genet* 14(21):3191-3201.
12. Eyre-Walker A, Woolfit M, & Phelps T (2006) The distribution of fitness effects of new deleterious amino acid mutations in humans. *Genetics* 173(2):891-900.
13. Boyko AR, *et al.* (2008) Assessing the evolutionary impact of amino acid mutations in the human genome. *PLoS Genet* 4(5):e1000083.
14. Mange AP (1964) Growth and inbreeding of a human isolate. *Hum Biol* 36:104-133.
15. Bleibtreu H (1964) Marriage and residence patterns in a genetic isolate Anthropology. Harvard, Cambridge).
16. Steinberg A, Bleibtreu H, Kurczynski T, Martin A, & Kurczynski E (1967) Genetic studies in an inbred human isolate. *Proceedings of the Third International Congress of Human Genetics*, eds Crow J & Neel J (Johns Hopkins University Press, Baltimore), pp 267-290.

17. Boycott KM, *et al.* (2010) A novel autosomal recessive malformation syndrome associated with developmental delay and distinctive facies maps to 16ptel in the Hutterite population. *Am J Med Genet A* 152A(6):1349-1356.
18. Çalışkan M, *et al.* (2011) Exome sequencing reveals a novel mutation for autosomal recessive non-syndromic mental retardation in the TECR gene on chromosome 19p13. *Hum Mol Genet* 20(7):1285-1289.
19. Huang L, *et al.* (2011) TMEM237 is mutated in individuals with a Joubert syndrome related disorder and expands the role of the TMEM family at the ciliary transition zone. *Am J Hum Genet* 89(6):713-730.
20. Bögershausen N, *et al.* (2013) Recessive TRAPPC11 mutations cause a disease spectrum of limb girdle muscular dystrophy and myopathy with movement disorder and intellectual disability. *Am J Hum Genet* 93(1):181-190.
21. Wiltshire KM, Hegele RA, Innes AM, & Brownell AK (2013) Homozygous lamin A/C familial lipodystrophy R482Q mutation in autosomal recessive Emery Dreifuss muscular dystrophy. *Neuromuscul Disord* 23(3):265-268.
22. Gerull B, *et al.* (2013) Homozygous founder mutation in desmocollin-2 (DSC2) causes arrhythmogenic cardiomyopathy in the Hutterite population. *Circ Cardiovasc Genet* 6(4):327-336.
23. Chong JX, Ouwenga R, Anderson RL, Waggoner DJ, & Ober C (2012) A population-based study of autosomal-recessive disease-causing mutations in a founder population. *Am J Hum Genet* 91(4):608-620.
24. Manatrinon S, Egger-Danner C, & Baumung R (2009) Estimating lethal allele frequencies in complex pedigrees via gene dropping approach using the example of Brown Swiss cattle. *Archiv Fur Tierzucht-Archives of Animal Breeding* 52(3):230-242.
25. Martin AO (1970) The founder effect in a human isolate: evolutionary implications. *American journal of physical anthropology* 32(3):351-367.
26. Keller L & Waller D (2002) Inbreeding effects in wild populations. *Trends in Ecology & Evolution* 17(5):230-241.
27. Overall AD, Ahmad M, & Nichols RA (2002) The effect of reproductive compensation on recessive disorders within consanguineous human populations. *Heredity (Edinb)* 88(6):474-479.
28. Ober C, Hyslop T, & Hauck WW (1999) Inbreeding effects on fertility in humans: evidence for reproductive compensation. *Am J Hum Genet* 64(1):225-231.
29. Hinckley JD, *et al.* (2013) Quantitative trait locus linkage analysis in a large Amish pedigree identifies novel candidate loci for erythrocyte traits. *Mol Genet Genomic Med* 1(3):131-141.
30. Macgregor S, *et al.* (2010) Legacy of mutiny on the Bounty: founder effect and admixture on Norfolk Island. *Eur J Hum Genet* 18(1):67-72.
31. Schull WJ, Nagano H, Yamamoto M, & Komatsu I (1970) The effects of parental consanguinity and inbreeding in Hirado, Japan. I. Stillbirths and prereproductive mortality. *Am J Hum Genet* 22(3):239-262.
32. Bittles AH & Black ML (2010) The impact of consanguinity on neonatal and infant health. *Early Hum Dev* 86(11):737-741.

33. Leridon H (1977) *Human Fertility: The Basic Components.* (University of Chicago Press, Chicago).
34. Mitchell KJ*, et al.* (2001) Functional analysis of secreted and transmembrane proteins critical to mouse development. *Nat Genet* 28(3):241-249.
35. White JK*, et al.* (2013) Genome-wide generation and systematic phenotyping of knockout mice reveals new roles for many genes. *Cell* 154(2):452-464.
36. Simons YB, Turchin MC, Pritchard JK, & Sella G (2014) The deleterious mutation load is insensitive to recent population history. *Nat Genet* 46(3):220-224.
37. Haldane JB (1935) The rate of spontaneous mutation of a human gene. *J Genet* 31:317-326.
38. Boone PM*, et al.* (2013) Deletions of recessive disease genes: CNV contribution to carrier states and disease-causing alleles. *Genome Res* 23(9):1383-1394.
39. Eyre-Walker A & Keightley PD (2007) The distribution of fitness effects of new mutations. *Nat Rev Genet* 8(8):610-618.
40. Dobzhansky T, Spassky B, & Tidwell T (1963) Genetics of natural populations. 32. Inbreeding and the mutational and balanced genetic loads in natural populations of Drosophila pseudoobscura. *Genetics* 48:361-373.
41. Malogolowkin-Cohen C, Levene H, Dobzhansky NP, & Simmons AS (1964) Inbreeding and the mutational and balanced loads in natural populations of *Drosophila willistoni. Genetics* 50:1299-1311.
42. McCune AR*, et al.* (2002) A low genomic number of recessive lethals in natural populations of bluefin killifish and zebrafish. *Science* 296(5577):2398-2401.
43. Halligan DL & Keightley PD (2003) How many lethal alleles? *Trends Genet* 19(2):57-59.
44. Gillespie JH (2004) *Population genetics : a concise guide* (Johns Hopkins University Press, Baltimore, Md.) 2nd Ed pp xiv, 214 p.
45. Ségurel L, Wyman M, & M. P (2014) Determinants of mutation rate variation in the human germline.  (Annual Reviews of Human Genetics).
46. Meyer WK*, et al.* (2012) Evaluating the evidence for transmission distortion in human pedigrees. *Genetics* 191(1):215-232.
47. Loucks C*, et al.* (2012) A shared founder mutation underlies restrictive dermopathy in Old Colony (Dutch-German) Mennonite and Hutterite patients in North America. *Am J Med Genet A* 158A(5):1229-1232.
48. Zielenski J*, et al.* (1993) Identification of the M1101K mutation in the cystic fibrosis transmembrane conductance regulator (CFTR) gene and complete detection of cystic fibrosis mutations in the Hutterite population. *Am J Hum Genet* 52(3):609-615.
49. Wolfram Research I (2012) Mathematica), 9.
50. Fujiwara TM*, et al.* (1989) Genealogical analysis of cystic fibrosis families and chromosome 7q RFLP haplotypes in the Hutterite Brethren. *Am J Hum Genet* 44(3):327-337.

**Figure Legends.**

**Figure. 1. The analysis pipeline for estimating the average number of recessive lethal mutations carried by humans. A.** A schematic diagram of the approach. The analysis procedures are described in black. Specific values and estimates for the Hutterites are provided in blue, including the point estimate of the mean number of mutations carried by a founder. **B.** The posterior distribution of the mean number of recessive lethal mutations carried by each haploid human genome, given the probability of manifestation and the number of diseases observed.

A.

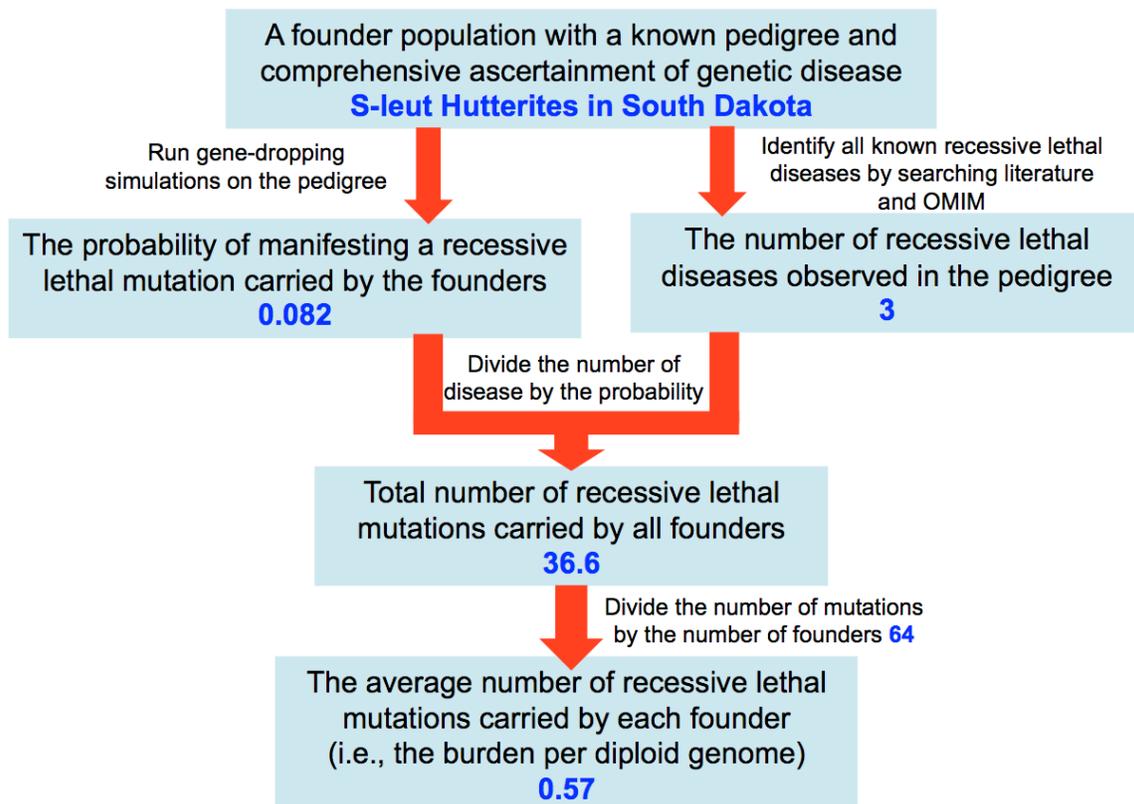

B.

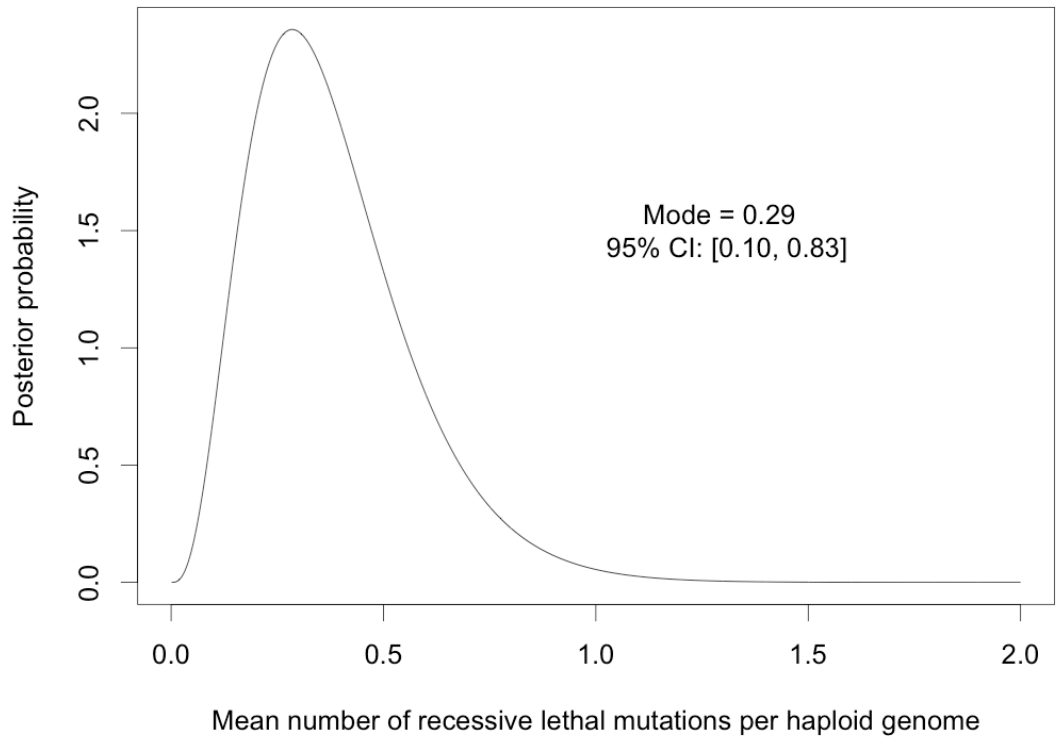

**SI Methods**

**I. The target size for mutations that lead to recessive lethal disorders between birth and reproductive age**

*i) The completely recessive case*

By taking the low mutation rate approximation and using the diffusion model, Kimura and Ohta (1) derived the expected sojourn time of a neutral new mutation in a finite, random-mating population at a given frequency $x$:

$$\tau(x) = \frac{2}{x}, \text{ for } \frac{1}{2N} \leq x \leq 1,$$

where $N$ is the effective population size (1).

Adopting Kimura and Ohta's model, Simons *et al.* (2) approximated the sojourn time of a recessive deleterious with selection coefficient $s$ in a finite, random-mating population by:

$$\tau(x) \approx \begin{cases} \frac{2}{x} & \text{if } \frac{1}{2N} \leq x \leq \frac{1}{\sqrt{2Ns}} \\ 0 & \text{if } \frac{1}{\sqrt{2Ns}} < x < 1 \end{cases}.$$

The sojourn time at frequency smaller than $\frac{1}{\sqrt{2Ns}}$ is the same as that of a neutral mutation, which is because most copies of the recessive mutation are concealed in heterozygotes, so the mutation behaves neutrally at that frequency range. However, when the frequency reaches $2Nx^2s \approx 1$, selection begins to act and thus pushes the allele frequency back to low level, so the recessive allele spends little time at frequency greater than $\frac{1}{\sqrt{2Ns}}$. The accuracy of this approximation is also confirmed by simulation (2).

For simplicity, we assume that there are *n* autosomal genes in the genome that each can lead to complete sterility or lethality between birth and reproductive age, and that gene *i* has $n_i$ functionally important sites that, once mutated, give rise to such recessive lethal alleles. We further assume that each site has approximated the same per generation mutation rate $\mu_{bp}$, so the total mutation rate to recessive lethal alleles at gene *i* is:

$$\mu_i = n_i \mu_{bp},$$

and the mean frequency of recessive lethal alleles at this gene *i* is:

$$\overline{q_i} = 2N\mu_i \cdot \int_{\frac{1}{2N}}^{\frac{1}{\sqrt{2N}}} x\tau(x)dx = 2N\mu_i \cdot \int_{\frac{1}{2N}}^{\frac{1}{\sqrt{2N}}} x \cdot \frac{2}{x} dx = 4N\mu_i(\frac{1}{\sqrt{2N}} - \frac{1}{2N}) = 2\mu_i(\sqrt{2N} - 1).$$

Therefore, the average total number of recessive lethal mutations carried by a haploid genome is:

$$\sum_{i=1}^{n} \overline{q_i} = \sum_{i=1}^{n} 2\mu_i(\sqrt{2N} - 1) = \sum_{i=1}^{n} 2n_i\mu_{bp}(\sqrt{2N} - 1) = 2\mu_{bp}(\sqrt{2N} - 1)\sum_{i=1}^{n} n_i,$$

where $\sum_{i=1}^{n} n_i$ is the target size for all autosomal recessive lethal mutations of interest. Assuming that the founders of S-leut Hutterites were drawn randomly from a population at equilibrium, each of them should have carried twice that number. Based on our estimate of 0.57 recessive lethal alleles per founder, a mutation rate of 1.2×10⁻⁸ per base pair per generation (3) and a diploid effective population size of 10,000, the target size is thus estimated to be 8.5×10⁴ base pairs.

*ii) The partially recessive case*

If a deleterious mutation leads to complete lethality (or sterility) in homozygotes ($s$=1) and a decrease of $hs$ in fitness of heterozygotes, selection against the deleterious allele would mainly come from the death of heterozygotes, because the death of homozygotes is a rare event (when the allele frequency is low). As in (i), we assume that there are $n$ autosomal genes in the genome that can mutate to such deleterious alleles, and that gene $i$ has $n_i$ such sites. Under these assumptions, the equilibrium frequency of the deleterious mutation at gene $i$ can be approximated as (4):

$$\bar{q}_i = \frac{\mu_i}{hs} = \frac{\mu_i}{h}.$$

Therefore, the total number of recessive lethal mutations carried by a random haploid genome is:

$$\sum_{i=1}^{n} \bar{q}_i = \sum_{i=1}^{n} \frac{\mu_i}{h} = \sum_{i=1}^{n} \frac{n_i \mu_{bp}}{h} = \frac{\mu_{bp}}{h} \sum_{i=1}^{n} n_i,$$

where $\sum_{i=1}^{n} n_i$ is the target size for all mutations that lead to lethality in homozygotes and a fitness decrease of $hs$ in heterozygotes. Assuming that $s$=1, $h$=0.01 and that the founders of S-leut Hutterites were drawn randomly from a population at equilibrium, each of them should have carried twice that number. Based on our estimate of 0.57 recessive lethal alleles per founder and a mutation rate of 1.2×10⁻⁸ per base pair per generation (3), the target size is thus estimated to be 2.4×10$^5$ base pairs.

**Tables S1: Three autosomal recessive lethal mutations found only in heterozygotes in the pedigree of S-leut Hutterites in South Dakota.**

| Name of Disease | OMIM Phenotype # | Gene | Mutation | Carrier freq. in Europeans | Carrier freq. in S-leut Hutterites | Reference |
|---|---|---|---|---|---|---|
| Bardet-Biedl syndrome | 209900 | *BBS2* | c.472-2A>G | Found only in Hutterites to date | 2.8% † | (5) |
| Dilated cardiomyopathy with ataxia syndrome | 610198 | *DNAJC19* | IVS3-1G>C | Found only in Hutterites to date | 2.8% † | (6) |
| Joubert syndrome /Meckel syndrome | 614424 | *TMEM237* | p.R18X | Found only in Hutterites to date | 8.0% † | (7, 8) |

† The allele frequencies are reported in Chong et al., 2012 (9).

**References**

1. Kimura M & Ohta T (1973) The age of a neutral mutant persisting in a finite population. *Genetics* 75(1):199-212.
2. Simons YB, Turchin MC, Pritchard JK, & Sella G (2014) The deleterious mutation load is insensitive to recent population history. *Nat Genet* 46(3):220-224.
3. Campbell CD*, et al.* (2012) Estimating the human mutation rate using autozygosity in a founder population. *Nat Genet* 44(11):1277-1281.
4. Haldane JB (1935) The rate of spontaneous mutation of a human gene. *J Genet* 31:317-326.
5. Innes AM*, et al.* (2010) A founder mutation in BBS2 is responsible for Bardet-Biedl syndrome in the Hutterite population: utility of SNP arrays in genetically heterogeneous disorders. *Clin Genet* 78(5):424-431.
6. Davey KM*, et al.* (2006) Mutation of DNAJC19, a human homologue of yeast inner mitochondrial membrane co-chaperones, causes DCMA syndrome, a novel autosomal recessive Barth syndrome-like condition. *J Med Genet* 43(5):385-393.
7. Schurig V, Bowen P, Harley F, & Schiff D (1980) The Meckel syndrome in the Hutterites. *Am J Med Genet* 5(4):373-381.
8. Huang L*, et al.* (2011) TMEM237 is mutated in individuals with a Joubert syndrome related disorder and expands the role of the TMEM family at the ciliary transition zone. *Am J Hum Genet* 89(6):713-730.
9. Chong JX, Ouwenga R, Anderson RL, Waggoner DJ, & Ober C (2012) A population-based study of autosomal-recessive disease-causing mutations in a founder population. *Am J Hum Genet* 91(4):608-620.